\documentclass[aps,prl,twocolumn,showpacs,superscriptaddress,groupedaddress]{revtex4-1}  
\usepackage{graphicx}  
\usepackage{dcolumn}   
\usepackage{bm}        
\usepackage{amssymb}   
\usepackage{amsmath}   
\usepackage{natbib}

\usepackage{color}

\begin{document}


\title{{Optimal chemotaxis in animal cell intermittent migration}}
\author{P. Romanczuk$^{1,2}$, G. Salbreux$^2$} 
\affiliation{$^1$ Department of Ecology and Evolutionary Biology, Princeton University, NJ 80544, USA, $^2$ Max Planck Institute for the Physics of Complex Systems,  N\"othnitzerstr. 38, 01187 Dresden, Germany}

\date{\today}
\begin{abstract}
Animal cells can sense chemical gradients without moving, and are faced with the challenge of migrating towards a target despite noisy information on the target position. Here we discuss optimal search strategies for a chaser that moves by switching between two phases of motion (ÒrunÓ and ÒtumbleÓ), reorienting itself towards the target during tumble phases, and performing a persistent random walk during run phases. We show that the chaser average run time can be adjusted to minimize the target catching time or the spatial dispersion of the chasers. We obtain analytical results for the catching time and for the spatial dispersion in the limits of small and large ratios of run time to tumble time, and scaling laws for the optimal run times. Our findings have implications for optimal chemotactic strategies in animal cell migration.
\end{abstract}
\maketitle



\begin{figure}
\begin{center}
\includegraphics[width=0.9\columnwidth]{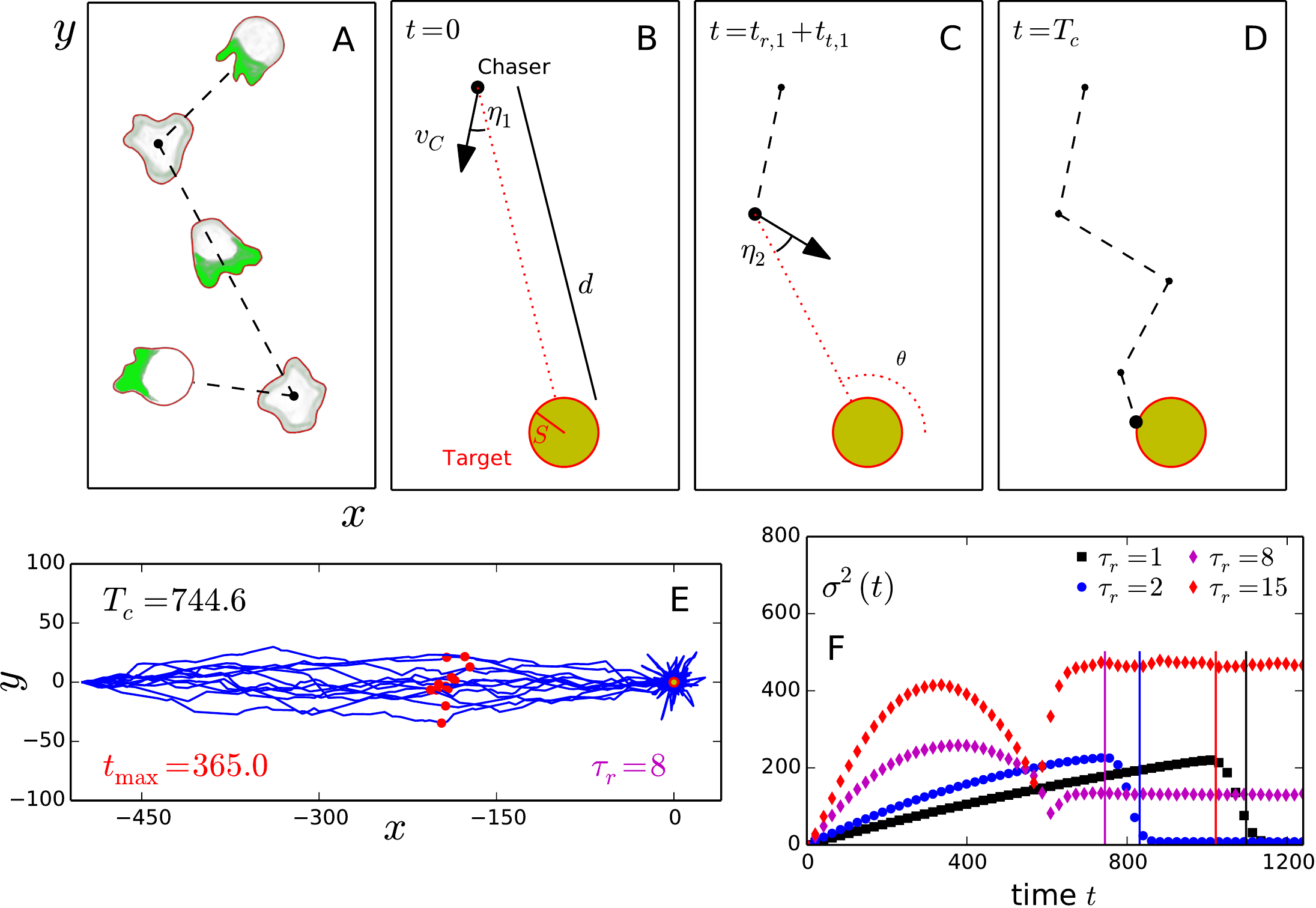}
\caption{{(A) Run and tumble migration of animal cells: Runs correspond to directed migration, of polarized cells with actin-rich protrusions (lamellipodia) at the leading edge (green). Tumbles correspond to reorientation events associated with repolarization events.} (B-D) Model schematic: A chasing particle aims at a target by switching between persistent motion towards the target (runs) and reorientation events (tumbles). After reorientation, the chaser orients itself towards the target with an error $\eta$. (E) Simulation snapshot of 20 chaser trajectories for $\tau_r=8$ and $t<T_c=744.6$; the red circles indicates chaser positions at maximal dispersion ($t_{\rm max}=388.0$). 
(F) Dispersion $\sigma^2(t)$ versus time for different run times. The vertical lines indicate the average catching time $T_c$. 
Parameters (d,e): $\theta_0=\pi$, $d_0=500$, $\epsilon=0.3$, $S=0.1$, $N=40000$. 
\label{fig:model_scheme}\label{fig:scheme_singlerun}}
\end{center}
\end{figure}

{Eukaryotic cells migration does not always occur continuously: bimodal motions with reorientation phases where cells  loose their polarity and phases of polarized, persistent motion have been reported in unicellular organisms \cite{polin_chlamydomonas_2009}, in cell migration occurring during development in the Zebrafish and Xenopus embryo  \cite{reichman-fried_autonomous_2004, minina_control_2007, blaser_migration_2006, theveneau_collective_2010}, or in mammalian cells \cite{potdar_bimodal_2009}. These two phases of motions have been denominated runs and tumbles \cite{reichman-fried_autonomous_2004} in analogy with {\it E. Coli} motion, although the corresponding migration mechanisms and chemical sensing of animal cells are very different to the ones employed by bacteria.  Possibly, cellular run and tumble behavior reflects the necessity for the cell to retract motile protrusions before forming new ones \cite{reichman-fried_autonomous_2004}: in that case, the run time is associated to the protrusion lifetime (cf. Fig \ref{fig:model_scheme}A).}

{Primordial germ (PG) cells in the Zebrafish embryo in particular have been reported to switch between two behavioural modes denoted ``run'' and ``tumbles'' \cite{reichman-fried_autonomous_2004}. In addition, PG cells move collectively in response to the chemokine SDF-1a towards the gonad during the first 24 hours of development \cite{reichman-fried_autonomous_2004}. In contrast to bacteria, after a tumble phase, PG cells choose a direction biased towards their target : animal cells are indeed capable of migration towards a source through sensing of chemical gradients \cite{swaney_eukaryotic_2010, levine_physics_2013}, allowing them to bias their motion towards a chemical source \cite{endres_accuracy_2008,ueda_stochastic_2007} without the need to sample the chemoattractant at different spatial positions.  }

{PG cells therefore display intermittent, directed migration towards a target. We investigate here the chemotactic efficiency of this class of motion with a minimal model. We ask in particular  (i) how does the time necessary to find the target depend on the properties of the cell motion, and (ii), how does the spatial dispersion of a group of cells evolve in time. Both questions are relevant for biological processes, where cells have to move to specific locations and maintain their cohesion \cite{doitsidou_guidance_2002}.}

In random search processes, a chaser without information on the target location performs a random walk until hitting the target by chance \cite{mendez_random_2014, benichou_optimal_2005}. In this context, the persistence length of a particle performing a random search can be optimised to minimise the search time \cite{tejedor_optimizing_2012}. In this work, we consider instead the optimal moving strategy for another class of motion, directed random search, where the chaser has noisy information on the location of the target, and needs to slow down to reorient.
More specifically, we consider an intermittent moving chaser which stops for a finite time to reorient towards its target (e.g. due to chemotaxis), with some detection error. This simple model accounts for the general situation where the reorientation of the chasing agent takes time, either due to physical constraints or to a finite detection time.

The main question is: how often should such a particle stop and reorient to efficiently move towards the target? Intuitively, short runs have the advantage of frequent reorientation towards the target, but at the cost of frequent stopping, while long runs may significantly deviate the particle from the location of the target, due to initial orientation errors and possible motion of the target. We show here that this results in an optimal run time which depends on the size and distance of the target and on the orientation errors that the chaser makes. Similarly, the effective diffusion of a collection of chasers can also be optimized, which might be essential for collective cell migration towards a common target.

We consider point-like agents (chasers) moving in two dimensions towards a disc-shaped target with radius $S$ located at position ${\bf x}_T$ (Fig. \ref{fig:scheme_singlerun}). The chaser position is denoted ${\bf x}_C$ and its direction of motion is given by the heading angle $\varphi$. The chasing agent switches between a run and a tumbling state. The tumbling phase lasts for a time $t_t$ during which the chaser does not move, after which the agent picks a direction ${\bf u}_C$ towards the target, with an additional angular ``detection'' error $\eta$ (Fig. \ref{fig:scheme_singlerun}). The initial polar angle of the chaser at the start of the run phase then reads 
\begin{equation}\label{eq:anglerun}
\varphi_i= \text{arg}({\bf x}_T - {\bf x}_C) + \eta.
\end{equation}
For simplicity, $\eta$ is drawn from a normal distribution with standard deviation $\epsilon$. In general, the detection error $\epsilon$ might depend on the distance to the target $d$, for instance due to the spatial variation of the chemoattractant. In this letter, we focus mainly on the simplest case of a constant $\epsilon$, and relax this assumption at the end of the letter.
During the run phase lasting for a time $t_r$, the particle is not able to further bias its motion towards the target, and moves with a constant speed $v_C$ and heading angle dynamics
$\dot \varphi = \eta_C(t)$, with
$\eta_C(t)$ a  Gaussian white noise with variance $D_{\varphi}$. For $D_{\varphi}=0$ the chaser moves on a straight line in the direction initially chosen $\varphi=\varphi_i$, whereas for $D_{\varphi}>0$ it performs a persistent random walk. 


 We assume here random switching between the run and reorientation phases to be given by a dichotomous Markov process. The durations of the different phases, $t_r$ and $t_t$, are exponentially distributed with an average run time $\tau_r$ and an average stopping time $\tau_t$.

To gain insight in the dynamics of the chasing motion, we performed simulations of the motion of a collection of chasers (Fig. \ref{fig:scheme_singlerun}). At time $t=0$ the chaser particles are placed at an angle $\theta_0$ and distance $d_0$ from the target placed at the origin, ${\bf x}_T=0$. Each chaser starts with a run with initial heading given by Eq. \ref{eq:anglerun}. Timescales are set by the tumbling time $\tau_t$, velocities are normalized by the chaser velocity $v_C$ and distances by $v_C \tau_t$.

To characterize the motion of chasers, we numerically evaluate the average catching time $T_c$ and the spatial dispersion in chaser positions,  $\sigma^2(t)=\langle(\mathbf{x}_C(t))^2\rangle-\langle\mathbf{x}_C(t)\rangle^2$. The dispersion exhibits different regimes over time (Fig. 1, \cite{supp_inf}): an initial diffusive spread, where $\sigma^2(t)$ increases roughly linearly, is followed by a transition to a stationary state, where $\sigma^2(t)$ assumes a constant value. 

\begin{figure}
\begin{center}
\includegraphics[width=0.9\columnwidth]{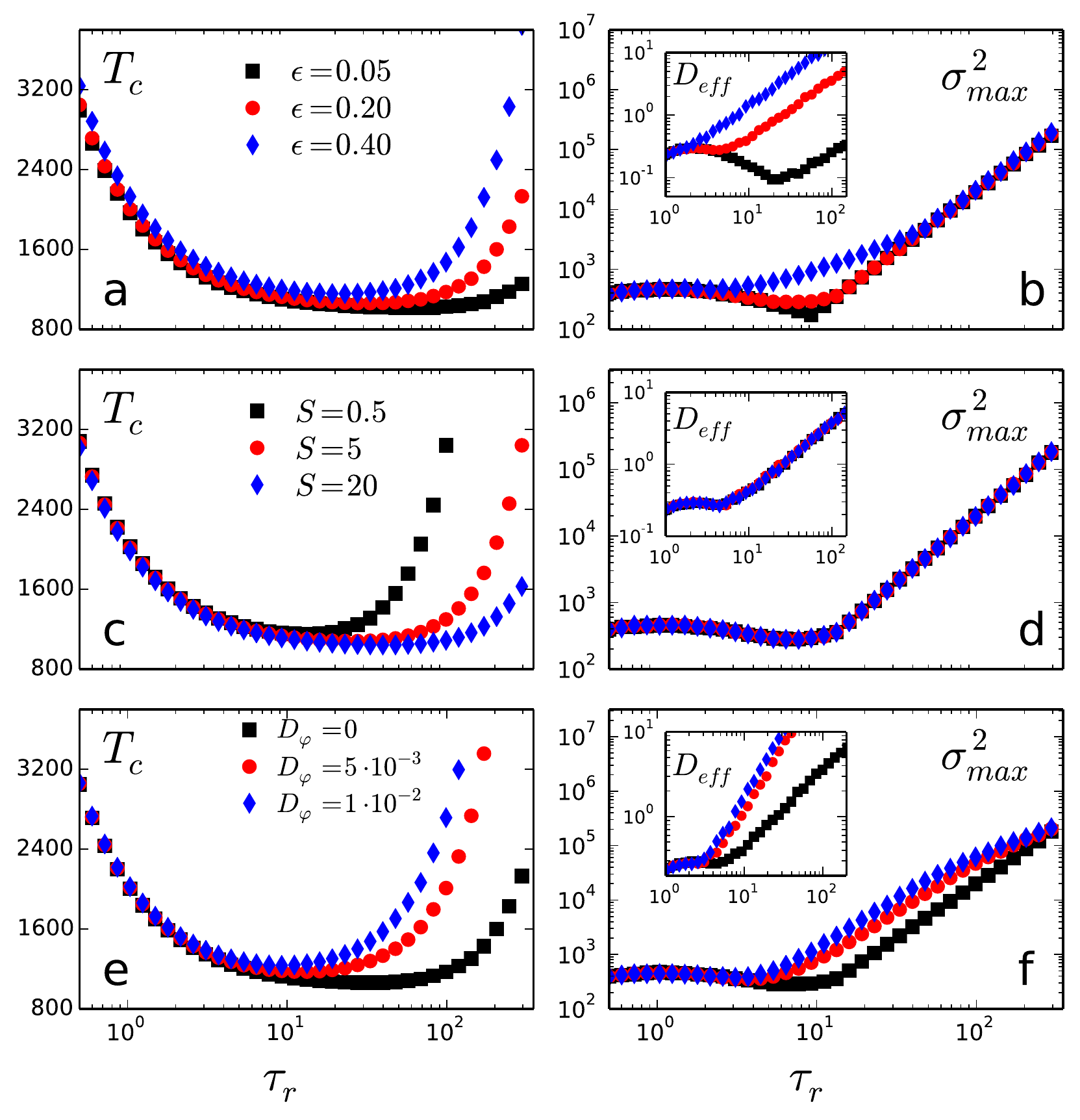}
\caption{Catching time $T_c$ (a,c,e) and maximum dispersion $\sigma^2_\text{max}$ (b,d,f) versus $\tau_r/\tau_t$ for different orientation errors $\epsilon$ (a,b), target radius $S$ (c,d) and angular noise $D_\varphi$ (e,f); insets show the effective diffusion coefficient $D_\text{eff}=\sigma^2_{max}/t_{max}$. Default parameters: $N=10^4$, $\theta_0=\pi$, $\epsilon=0.2$, $d_0=1000$, $\tau_t=1$, $v_C=1$, $S=1$, $D_\varphi=0$.\label{fig:simulation}}
\end{center}
\end{figure}

We then evaluated the average catching time $T_c$,  the maximum dispersion during the catching process $\sigma_{\max}^2$, and the initial effective diffusion $D_{\rm{eff}}=\sigma^2_{\max}/t_{\max}$ with $t_{\max}$ being the time of maximum dispersion. When varying the ratio of run to tumble time $\tau_r/\tau_t$, we observe that $T_c$ is minimal for an optimal value of $\tau_r/\tau_t$ (Fig. \ref{fig:simulation}). $\sigma^2_{\max}$ also exhibits a minimum for zero or a finite value of the run time $\tau_r$.
This behavior can be understood intuitively as follows: for short run times $\tau_r$, the chaser reorients frequently, allowing it to follow accurately the target; the frequent reorientations however slow the chaser down. 
For long run times on the other end, the chaser has a high probability to miss the target due to orientation errors, leading to large displacements away from the target. 
When varying the rotational diffusion of the chaser $D_{\varphi}$, we find that for $\tau_r>1/D_{\varphi}$ the chaser ``forgets'' its initial direction during each run and the catching time sharply increases, while for short run times the catching time and chaser dispersion are unchanged  (Fig. \ref{fig:simulation}). Because rotational diffusion then essentially introduces a maximum value of $\tau_r$ above which the catching process becomes undirected, we focus on the limit $D_{\varphi}\rightarrow 0$ in the following.

\begin{figure}
\begin{center}
\includegraphics[width=\columnwidth]{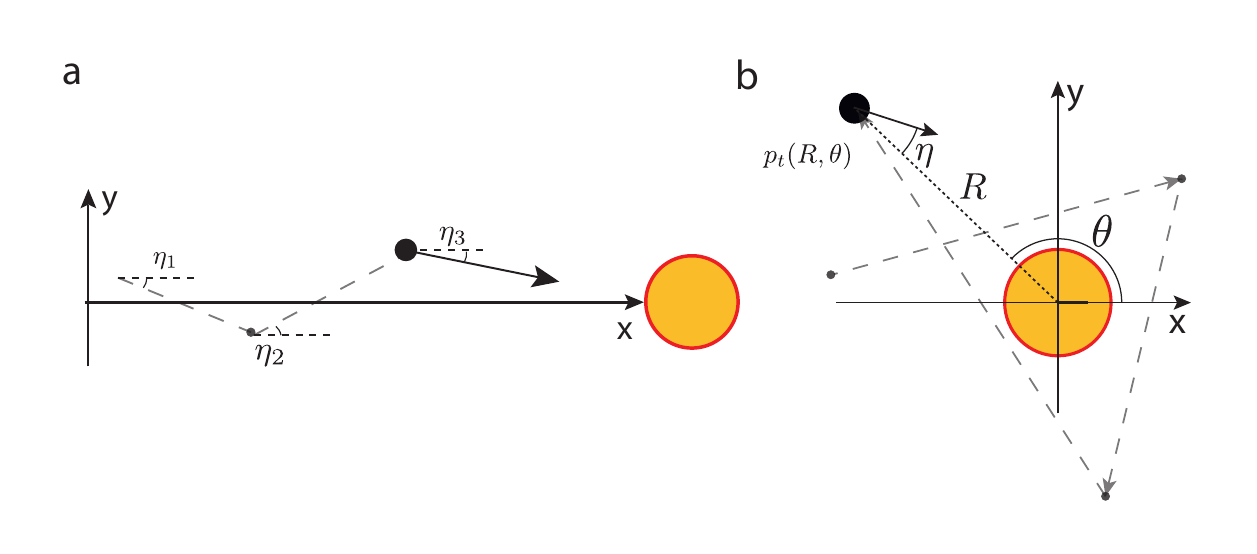}
\caption{Limit cases of the chaser motion. (a) For small enough runs and away from the target, the chaser undergoes a nearly one-dimensional motion. (b) For large runs and close to the target, the chaser moves around the target and has a fixed probability of hitting the target at each run. \label{fig:schemelimitcases}}
\end{center}
\end{figure}

To understand the origin of the optimal behavior, we note that two regimes emerge from the analysis of numerical simulations depending on the ratio between the average run length and the distance to the target (Fig. \ref{fig:schemelimitcases}). For $v_c \tau_r\ll |{\bf x}_C|$ (large distances and short runs), the chasing motion is dominated by a slow effective drift towards the target. For $v_c\tau_r \gtrsim |{\bf x}_C|$ (small distances and large runs), the chasers approach the vicinity of the target and the catching process is primarily dominated by the probability of hitting the target within a single run. The total catching time is then the sum of the time to approach the target (first regime), and the time to catch it (second regime). A visualization of the catching process for different average run times is provided in the Supplementary Material \cite{supp_inf}.

Based on these observations, we start by considering the limit of short runs with chasers positioned far from the target with the initial distance $d_0\gg S$.  We assume here, without loss of generality due to rotational symmetry, that the chaser initial position is along the x axis, ${\bf x}_c(t=0)=-d_{0} \mathbf{e}_x$. The chaser then follows a nearly one-dimensional motion along the x-direction, such that the desired chaser heading angle is small ($\varphi\approx0$).  
The system can be described in terms of the spatial probability density functions (PDFs) for chaser particles in the run and tumble phase $p_{r}({\bf x},\varphi)$ and $p_{t}({\bf x})$ \cite{othmer_models_1988}, where the probability distribution is conditioned to the initial position and angle $(\mathbf{x}_0,\varphi_0)$. The general evolution equations for the PDFs in the frame of the target read
\begin{align}
\partial_t p_r(\mathbf{x},\varphi)=& -{\bf u}_r.\boldsymbol{\nabla} p_r(\mathbf{x},\varphi) - \frac{1}{\tau_r} p_r(\mathbf{x},\varphi)+\frac{1}{\tau_t} T(\varphi) p_t(\mathbf{x}) \quad \label{eq:eom_pr}\\ 
\partial_t p_t (\mathbf{x}) =& - \frac{1}{\tau_t} p_t (\mathbf{x})+\frac{1}{\tau_r}\int_0^{2\pi} d\varphi  p_r (\mathbf{x},\varphi)\label{eq:eom_ps}
\end{align}
where ${\bf u}_r =v_C {\bf u}_C(\varphi)$ is the drift velocity in the run phase, $T(\varphi)=\frac{1}{\sqrt{2\pi}\epsilon}\text{exp}\left( -\frac{\varphi^2}{2\epsilon^2}\right)$ is the probability of reorientation in the $\varphi$ direction after a tumble, $\boldsymbol{\nabla}$ denotes the gradient operator acting on the $\mathbf{x}$ dependency of $p_r(\mathbf{x},\varphi)$, and the time dependency of $p_r$ and $p_t$ is implicitly implied.
By deriving moment equations from Eqs. \ref{eq:eom_pr}-\ref{eq:eom_ps} (Supp. Mat), we find that the average chaser velocity towards the target relaxes in a time $\tau_r \tau_t/(\tau_r+\tau_t)$ to the stationary value
\begin{eqnarray}
\langle v \rangle=\frac{v_C\tau_r e^{-\frac{\epsilon^2}{2}}}{\tau_r+\tau_t}, 
\end{eqnarray}
while the chaser dispersion $\sigma^2(t)=D_{\text{eff}} t$ is predicted to increase linearly in time with the effective diffusion coefficient
\begin{eqnarray}
D_{\text{eff}}  = \frac{2 v_C^2\tau_r^2}{\tau_r+\tau_t}\left[1-(1-\frac{\tau_t^2}{(\tau_r+\tau_t)^2})e^{-\epsilon^2}\right] .
\end{eqnarray}
For $\epsilon=0$, $D_\text{eff}=\frac{2 v_C^2\tau_r^2\tau_t^2}{(\tau_r+\tau_t)^3}$: in that limit, the diffusion is only due to the stochasticity in run and tumble times, and for large run time the diffusion constant decreases with $\tau_r$, since each additional run introduces dispersion by generating different run lengths between different chasers (a zero run time also minimizes the diffusion, but the catching time then diverges). For $\epsilon\neq 0$, there is an optimal run time minimizing the diffusion coefficient, given for $\epsilon\rightarrow 0$ by $\tau_r/\tau_t=1/\epsilon$: large runs indeed increase dispersion, because they tend to amplify initial errors in the heading angle.

The average approach time to the target is given in that drift-dominated limit by
\begin{equation}
\label{Tc1D}
T_c^{(1)}=\frac{d_0-S}{v_C \frac{ \tau_r}{\tau_r+\tau_t} e^{-\frac{\epsilon^2}{2}}},
\end{equation}
which agrees with numerical simulations (Fig. S1). Therefore, in the limit of small runs, increasing the run time is always favorable.  

We now discuss the opposite limit when the target is sufficiently close to the chaser to be reached within a single run (Fig. \ref{fig:schemelimitcases}B). 
We assume that the probability distribution of the chaser position at tumble, $p_t$, has relaxed to a steady state. The steady-state distribution is then isotropic. At lowest order in the orientation error $\epsilon$, the chaser-target distance $R=|{\bf x}_c|$ is distributed according to an exponential distribution, $p_t(R)=\frac{1}{\tau_r v_c}e^{-\frac{R}{\tau_r v_c}}$, with the corresponding standard deviation of chaser position $\sigma_{\rm{stat}}=\tau_r v_c$.
To evaluate the catching time, we note that the probability of hitting the target in one run, after each tumble, is given by
\begin{equation}
p_{hit}=\int_{R=0}^{\infty}dR p_t(R)p_\text{dir}(R) p_\text{reach}(R)
\end{equation}
where $p_{\text{dir }}(R)$ is the probability of choosing a direction towards the target, and $p_\text{reach}(R)$ is the probability of performing a sufficiently long run to actually reach the target. We start by evaluating $p_\text{dir}$. 
The chaser is heading towards the target provided that the detection error is sufficiently small, $|\eta|<\frac{S}{R}$.
As a result, the probability of choosing the right direction by a favorable choice of error $\eta$ is given by
\begin{equation}
p_\text{dir}(R)=2\int_{0}^{\frac{ S}{R\sqrt{2}\epsilon}}
%
dx\frac{ e^{-x^2}}{\sqrt{\pi}}
\end{equation}
To obtain $p_\text{reach}(R)$, we note that (1) at lowest order in $\frac{S}{R}$, the run time has to to be larger than $t_\text{catch} =R/v_C$ to reach the target, and (2) run times are taken out of an exponential distribution; therefore, $p_\text{reach}=e^{-R/\tau_r v_C}$.
 \begin{figure}
\begin{center}
\includegraphics[width=\columnwidth]{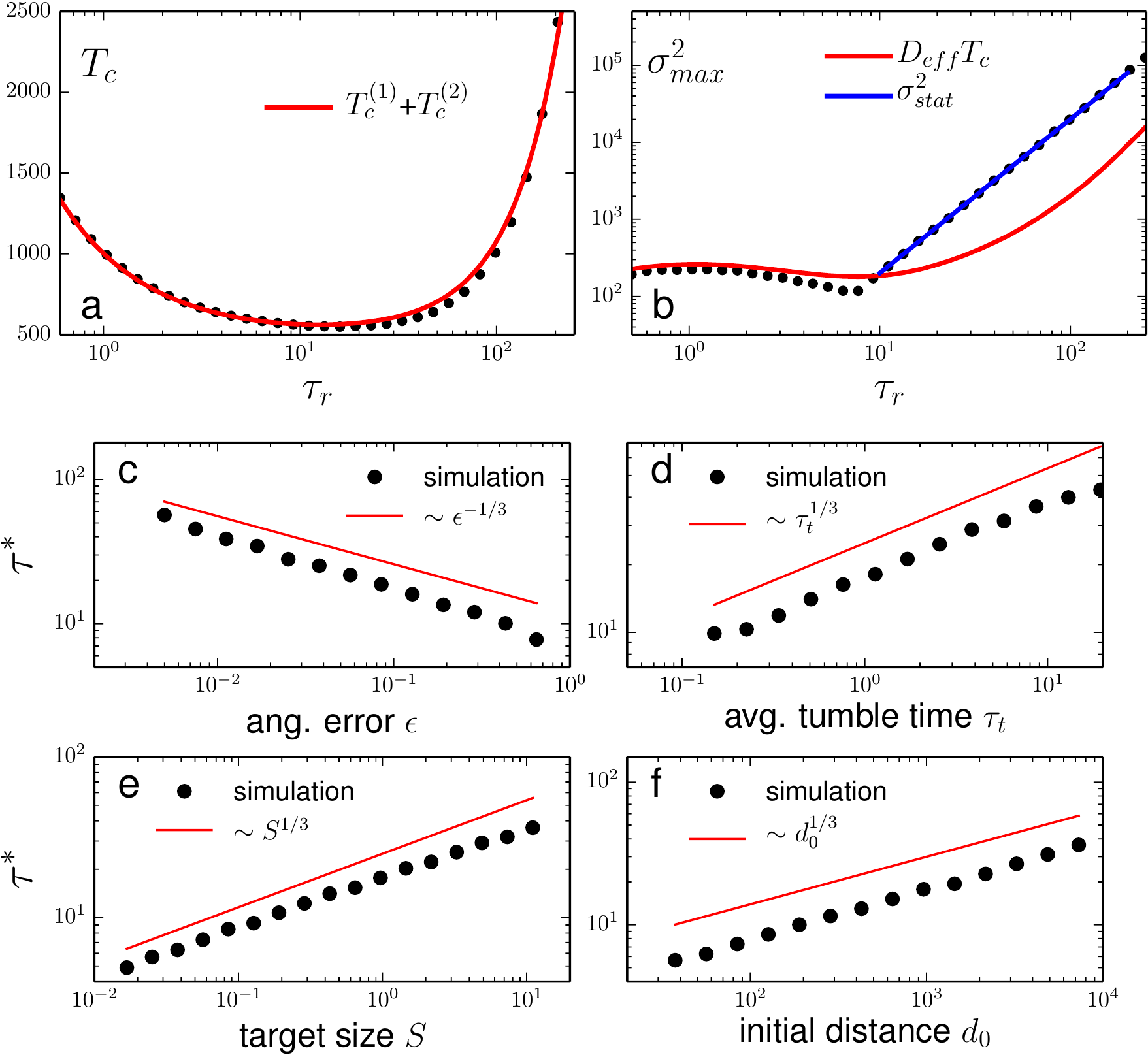}

\caption{
Comparison of simulation results and theory: (a) $T_c$ and analytical prediction $T_c=T_c^{(1)}+T_c^{2}$; (b) $\sigma^2_\text{max}$ with $D_\text{eff}T_c$ for the ``quasi-1D'' approach and $\sigma_\text{stat}^2=(\tau_r v_c)^2$ for the ``single run'' encounter (right). Parameters: $N=10^4$, $\epsilon=0.1$,  $S=1$, $d_{0}=500$. 
Scaling of the optimal run time $\tau_r^*$ minimizing $T_c$, (c) with the angular error $\epsilon$, (d) the average tumble time $\tau_t$, (e) the target size $S$ and (f) the initial distance $d$. The solid lines indicate the predicted dependence from theory. Simulation parameters: $N=10^4$, $v_T=0.1$, and unless parameters are varied, $\tau_t=1$, $v_C=1$, $\epsilon=0.2$, $S=1$, $d_0=500$ and $\theta_0=\pi$.
\label{fig:comparison_sim_theo}
}
\end{center}
\end{figure}

Overall, we then obtain $p_\text{hit}\simeq \frac{1}{2} M(\frac{\sqrt{2} S}{\tau_r v_C  \epsilon})$ with $M(x)=\int_0^{\infty} dy\, e^{-y}\text{Erf}(x/y)$ \cite{Note3},
and the average catching time from the vicinity of the target reads
\begin{eqnarray}
\label{Tcsinglerun}
T_c^{(2)}\simeq  \frac{2(\tau_r+\tau_t)}{M(\frac{\sqrt{2}S}{ \tau_r v_C  \epsilon})}.
\end{eqnarray}
Therefore, in the limit of large runs, decreasing the run time is favorable, an opposite behavior to the short run time limit discussed before.

The total time to reach the target, $T_c$, is the sum of the time required to approach the target (Eq. \ref{Tc1D}) and the time to catch it in its vicinity (Eq. \ref{Tcsinglerun}), $T_c=T_c^{(1)}+T_c^{(2)}$. The resulting analytic prediction accounts very well for the catching time obtained by simulations (Fig. \ref{fig:comparison_sim_theo}a).  To estimate the optimal run time, we look for the minimum of the total catching time $T_c$. We find that for intermediate orientation error $\frac{S}{\sqrt{d v_c\tau_t}} \ll \epsilon \ll 1$, the optimal run time $\tau_r^*$ scales as \cite{supp_inf}
\begin{equation}
\label{ScalingOptimalRunTime}
\tau_r^*\sim\left(\frac{dS}{v_c^2 \epsilon} \tau_t \right)^{1/3}
\end{equation}
The condition can be rewritten as an optimal run length $\tau_r^* v_C\sim (dS \tau_t v_c/\epsilon)^{\frac{1}{3}}$, scaling with the geometric mean of the three lengths of the problem. The predicted scaling is verified by numerical simulations  (Fig. \ref{fig:comparison_sim_theo}c-f). The intermittent directed chaser has an optimal run length strikingly different from the optimal persistence length of a random searcher, which was found to be of the order of the system size \cite{tejedor_optimizing_2012}. It further implies that performing short length runs is better for small target sizes, when the distance to the target is close, and for short tumbles. The optimal run length is also crucially dependent on the orientation error: quite naturally, smaller orientation errors allow to perform larger runs.
\begin{figure}
\begin{center}
\includegraphics[width=0.95\columnwidth]{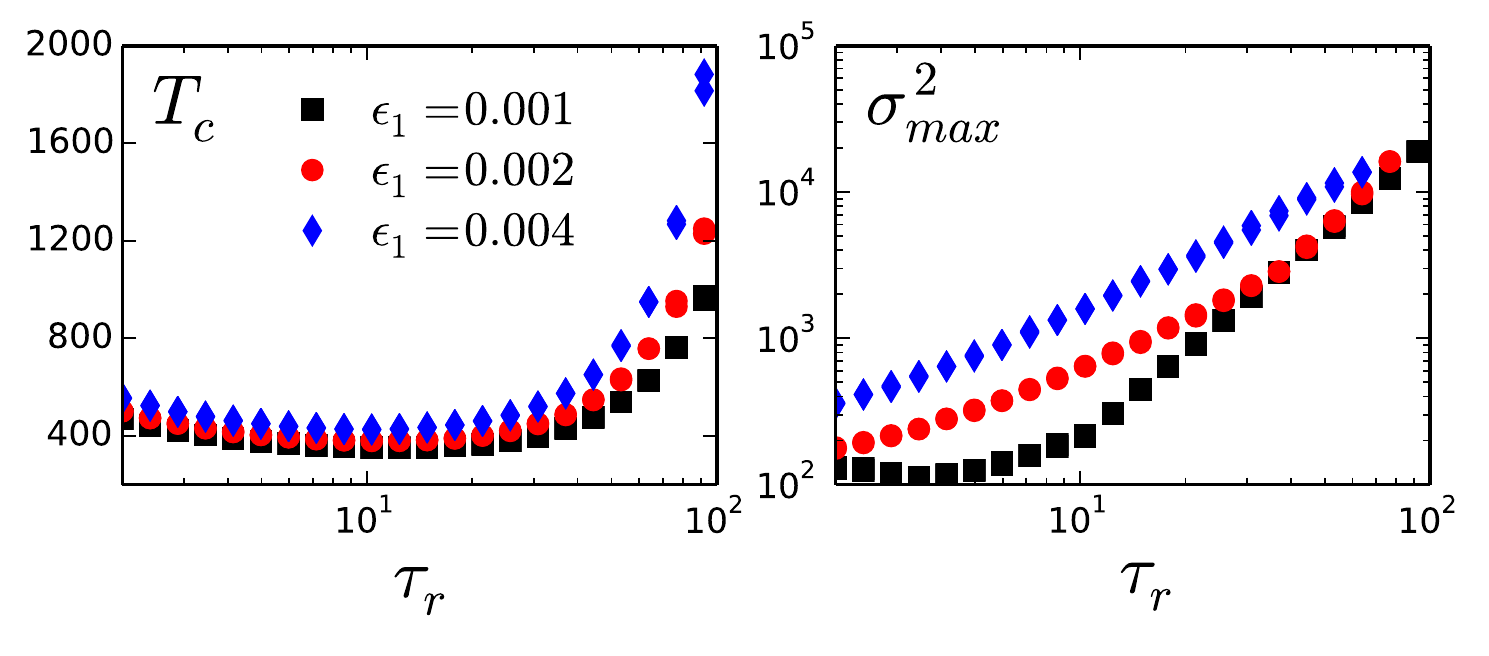}
\caption{
Catching time $T_c$ and maximum dispersion $\sigma^2_\text{max}$ versus $\epsilon_1$. Simulation parameters: $N=10^4$, $\tau_t=1$, $v_C=1$, $\epsilon_0=0.1$, $S=1$, $d_0=300$ and $\theta_0=\pi$. 
\label{fig:sim_epsilon1}
}
\end{center}
\end{figure}

In the minimal model discussed so far, we have assumed that the reorientation error $\epsilon$ is independent of the distance to the target $d(t)=|\mathbf{x_c}(t)|$. For directional sensing however, the detection accuracy of the target direction is likely to decrease with $d(t)$. To account for this, we numerically investigated an extension of our model by assuming that an arbitrary error function $\epsilon(d)$ can be approximated by a linear function $\epsilon(d)\approx\epsilon_0+\epsilon_1 d$ with $\epsilon_1>0$. Simulations results for a finite $\epsilon_1$ show the same qualitative behavior as the minimal model (Fig. \ref{fig:sim_epsilon1}): an optimal run time minimizing the catching time still exists. When increasing $\epsilon_1$, the optimal run time minimizing the dispersion moves towards zero, similar to increasing the overall constant error $\epsilon$ (Fig. \ref{fig:simulation}b).

We have also considered the situation where the target moves balistically with a finite, but small, speed $v_T<v_C$, and found that the same qualitative findings hold \cite{supp_inf}. A crucial difference for a finite target speed $v_T$ is the existence of a regime where the catching time $T_c$ diverges, occurring above a threshold detection error $\epsilon$ or for short run time $\tau_r$ \cite{supp_inf}.

The robustness of our qualitative results to these modifications suggests that the minimal model introduced here exhibits physical properties which will be preserved for more realistic description of cells moving with intermittent motion. Developmental processes require cell migration occurring at the scale of the organism \cite{richardson_mechanisms_2010}: it would be interesting to experimentally test whether migrating cells operate near the optimum value predicted by Eq. \ref{ScalingOptimalRunTime} in this context.
{Recently, Minina {\em et al.} \cite{minina_control_2007} have analyzed in detail guided migration of progenitor cell in vivo. Based on their experimental results they suggest that progenitor cells might adaptively decrease their run length by downregulating receptor signaling close to the source of the relevant chemokine. A disruption of this control mechanism for run length was shown to hinder precise arrival of cells at the target. These empirical findings are confirmed by our theoretical work, in particular by the predicted decrease in optimal run length with decreasing distance to the target (Fig. \ref{fig:comparison_sim_theo}f). This indeed suggests that PG cells have evolved complex chemotactic strategies for precise spatial and temporal arrival at target sites, to ensure successful development of the embryo. We expect that further research combining new quantitative experiments and more elaborate theoretical models will provide important insights into the mechanism and function of directed cell migration in living systems.}

{From a more general point of view with respect to theory of search processes, our simple but generic model can be viewed as an investigation of optimal directed search, which is fundamentally different from the random search processes predominantly studied up to date. 
}  
 


{\it Acknowledgements} We thank Ewa Paluch for bringing cellular run and tumble motion to our attention and for stimulating discussions, and Vasily Zaburdaev for helpful comments on the manuscript. PR gratefully acknowledges the hospitality of the Kavli Institute for Theoretical Physics (UCSB). This research was supported in part by the National Science Foundation under Grant No. NSF PHY11-25915.
\bibliographystyle{Myapsrev}
\bibliography{Manuscript}

\begin{thebibliography}{18}
\expandafter\ifx\csname natexlab\endcsname\relax\def\natexlab#1{#1}\fi
\expandafter\ifx\csname bibnamefont\endcsname\relax
  \def\bibnamefont#1{#1}\fi
\expandafter\ifx\csname bibfnamefont\endcsname\relax
  \def\bibfnamefont#1{#1}\fi
\expandafter\ifx\csname citenamefont\endcsname\relax
  \def\citenamefont#1{#1}\fi
\expandafter\ifx\csname url\endcsname\relax
  \def\url#1{\texttt{#1}}\fi
\expandafter\ifx\csname urlprefix\endcsname\relax\def\urlprefix{URL }\fi
\providecommand{\bibinfo}[2]{#2}
\providecommand{\eprint}[2][]{\url{#2}}

\bibitem[{\citenamefont{Polin et~al.}(2009)\citenamefont{Polin, Tuval,
  Drescher, Gollub, and Goldstein}}]{polin_chlamydomonas_2009}
\bibinfo{author}{\bibfnamefont{M.}~\bibnamefont{Polin}},
  \bibinfo{author}{\bibfnamefont{I.}~\bibnamefont{Tuval}},
  \bibinfo{author}{\bibfnamefont{K.}~\bibnamefont{Drescher}},
  \bibinfo{author}{\bibfnamefont{J.~P.} \bibnamefont{Gollub}},
  \bibnamefont{and} \bibinfo{author}{\bibfnamefont{R.~E.}
  \bibnamefont{Goldstein}}, \bibinfo{journal}{Science}
  \textbf{\bibinfo{volume}{325}}, \bibinfo{pages}{487} (\bibinfo{year}{2009}),
  \bibinfo{note}{{PMID:} 19628868}.

\bibitem[{\citenamefont{Reichman-Fried
  et~al.}(2004)\citenamefont{Reichman-Fried, Minina, and
  Raz}}]{reichman-fried_autonomous_2004}
\bibinfo{author}{\bibfnamefont{M.}~\bibnamefont{Reichman-Fried}},
  \bibinfo{author}{\bibfnamefont{S.}~\bibnamefont{Minina}}, \bibnamefont{and}
  \bibinfo{author}{\bibfnamefont{E.}~\bibnamefont{Raz}},
  \bibinfo{journal}{Developmental Cell} \textbf{\bibinfo{volume}{6}},
  \bibinfo{pages}{589} (\bibinfo{year}{2004}).

\bibitem[{\citenamefont{Minina et~al.}(2007)\citenamefont{Minina,
  Reichman-Fried, and Raz}}]{minina_control_2007}
\bibinfo{author}{\bibfnamefont{S.}~\bibnamefont{Minina}},
  \bibinfo{author}{\bibfnamefont{M.}~\bibnamefont{Reichman-Fried}},
  \bibnamefont{and} \bibinfo{author}{\bibfnamefont{E.}~\bibnamefont{Raz}},
  \bibinfo{journal}{Current Biology} \textbf{\bibinfo{volume}{17}},
  \bibinfo{pages}{1164} (\bibinfo{year}{2007}).

\bibitem[{\citenamefont{Blaser et~al.}(2006)\citenamefont{Blaser,
  Reichman-Fried, Castanon, Dumstrei, Marlow, Kawakami, Solnica-Krezel,
  Heisenberg, and Raz}}]{blaser_migration_2006}
\bibinfo{author}{\bibfnamefont{H.}~\bibnamefont{Blaser}},
  \bibinfo{author}{\bibfnamefont{M.}~\bibnamefont{Reichman-Fried}},
  \bibinfo{author}{\bibfnamefont{I.}~\bibnamefont{Castanon}},
  \bibinfo{author}{\bibfnamefont{K.}~\bibnamefont{Dumstrei}},
  \bibinfo{author}{\bibfnamefont{F.~L.} \bibnamefont{Marlow}},
  \bibinfo{author}{\bibfnamefont{K.}~\bibnamefont{Kawakami}},
  \bibinfo{author}{\bibfnamefont{L.}~\bibnamefont{Solnica-Krezel}},
  \bibinfo{author}{\bibfnamefont{C.-P.} \bibnamefont{Heisenberg}},
  \bibnamefont{and} \bibinfo{author}{\bibfnamefont{E.}~\bibnamefont{Raz}},
  \bibinfo{journal}{Developmental Cell} \textbf{\bibinfo{volume}{11}},
  \bibinfo{pages}{613} (\bibinfo{year}{2006}).

\bibitem[{\citenamefont{Theveneau et~al.}(2010)\citenamefont{Theveneau,
  Marchant, Kuriyama, Gull, Moepps, Parsons, and
  Mayor}}]{theveneau_collective_2010}
\bibinfo{author}{\bibfnamefont{E.}~\bibnamefont{Theveneau}},
  \bibinfo{author}{\bibfnamefont{L.}~\bibnamefont{Marchant}},
  \bibinfo{author}{\bibfnamefont{S.}~\bibnamefont{Kuriyama}},
  \bibinfo{author}{\bibfnamefont{M.}~\bibnamefont{Gull}},
  \bibinfo{author}{\bibfnamefont{B.}~\bibnamefont{Moepps}},
  \bibinfo{author}{\bibfnamefont{M.}~\bibnamefont{Parsons}}, \bibnamefont{and}
  \bibinfo{author}{\bibfnamefont{R.}~\bibnamefont{Mayor}},
  \bibinfo{journal}{Developmental Cell} \textbf{\bibinfo{volume}{19}},
  \bibinfo{pages}{39} (\bibinfo{year}{2010}), \bibinfo{note}{{PMID:} 20643349}.

\bibitem[{\citenamefont{Potdar et~al.}(2009)\citenamefont{Potdar, Lu, Jeon,
  Weaver, and Cummings}}]{potdar_bimodal_2009}
\bibinfo{author}{\bibfnamefont{A.~A.} \bibnamefont{Potdar}},
  \bibinfo{author}{\bibfnamefont{J.}~\bibnamefont{Lu}},
  \bibinfo{author}{\bibfnamefont{J.}~\bibnamefont{Jeon}},
  \bibinfo{author}{\bibfnamefont{A.~M.} \bibnamefont{Weaver}},
  \bibnamefont{and} \bibinfo{author}{\bibfnamefont{P.~T.}
  \bibnamefont{Cummings}}, \bibinfo{journal}{Annals of biomedical engineering}
  \textbf{\bibinfo{volume}{37}}, \bibinfo{pages}{230} (\bibinfo{year}{2009}),
  \bibinfo{note}{{PMID:} 18982450 {PMCID:} {PMC3586332}}.

\bibitem[{\citenamefont{Swaney et~al.}(2010)\citenamefont{Swaney, Huang, and
  Devreotes}}]{swaney_eukaryotic_2010}
\bibinfo{author}{\bibfnamefont{K.~F.} \bibnamefont{Swaney}},
  \bibinfo{author}{\bibfnamefont{C.-H.} \bibnamefont{Huang}}, \bibnamefont{and}
  \bibinfo{author}{\bibfnamefont{P.~N.} \bibnamefont{Devreotes}},
  \bibinfo{journal}{Annual Review of Biophysics} \textbf{\bibinfo{volume}{39}},
  \bibinfo{pages}{265} (\bibinfo{year}{2010}), \bibinfo{note}{{PMID:}
  20192768}.

\bibitem[{\citenamefont{Levine and Rappel}(2013)}]{levine_physics_2013}
\bibinfo{author}{\bibfnamefont{H.}~\bibnamefont{Levine}} \bibnamefont{and}
  \bibinfo{author}{\bibfnamefont{W.-J.} \bibnamefont{Rappel}},
  \bibinfo{journal}{Physics Today} \textbf{\bibinfo{volume}{66}},
  \bibinfo{pages}{24} (\bibinfo{year}{2013}).

\bibitem[{\citenamefont{Endres and Wingreen}(2008)}]{endres_accuracy_2008}
\bibinfo{author}{\bibfnamefont{R.~G.} \bibnamefont{Endres}} \bibnamefont{and}
  \bibinfo{author}{\bibfnamefont{N.~S.} \bibnamefont{Wingreen}},
  \bibinfo{journal}{Proceedings of the National Academy of Sciences}
  \textbf{\bibinfo{volume}{105}}, \bibinfo{pages}{15749}
  (\bibinfo{year}{2008}).

\bibitem[{\citenamefont{Ueda and Shibata}(2007)}]{ueda_stochastic_2007}
\bibinfo{author}{\bibfnamefont{M.}~\bibnamefont{Ueda}} \bibnamefont{and}
  \bibinfo{author}{\bibfnamefont{T.}~\bibnamefont{Shibata}},
  \bibinfo{journal}{Biophysical Journal} \textbf{\bibinfo{volume}{93}},
  \bibinfo{pages}{11} (\bibinfo{year}{2007}).

\bibitem[{\citenamefont{Doitsidou et~al.}(2002)\citenamefont{Doitsidou,
  Reichman-Fried, Stebler, Köprunner, Dörries, Meyer, Esguerra, Leung, and
  Raz}}]{doitsidou_guidance_2002}
\bibinfo{author}{\bibfnamefont{M.}~\bibnamefont{Doitsidou}},
  \bibinfo{author}{\bibfnamefont{M.}~\bibnamefont{Reichman-Fried}},
  \bibinfo{author}{\bibfnamefont{J.}~\bibnamefont{Stebler}},
  \bibinfo{author}{\bibfnamefont{M.}~\bibnamefont{Köprunner}},
  \bibinfo{author}{\bibfnamefont{J.}~\bibnamefont{Dörries}},
  \bibinfo{author}{\bibfnamefont{D.}~\bibnamefont{Meyer}},
  \bibinfo{author}{\bibfnamefont{C.~V.} \bibnamefont{Esguerra}},
  \bibinfo{author}{\bibfnamefont{T.}~\bibnamefont{Leung}}, \bibnamefont{and}
  \bibinfo{author}{\bibfnamefont{E.}~\bibnamefont{Raz}},
  \bibinfo{journal}{Cell} \textbf{\bibinfo{volume}{111}}, \bibinfo{pages}{647}
  (\bibinfo{year}{2002}).

\bibitem[{\citenamefont{M{\'e}ndez et~al.}(2014)\citenamefont{M{\'e}ndez,
  Campos, and Bartumeus}}]{mendez_random_2014}
\bibinfo{author}{\bibfnamefont{V.}~\bibnamefont{M{\'e}ndez}},
  \bibinfo{author}{\bibfnamefont{D.}~\bibnamefont{Campos}}, \bibnamefont{and}
  \bibinfo{author}{\bibfnamefont{F.}~\bibnamefont{Bartumeus}}, in
  \emph{\bibinfo{booktitle}{Stochastic Foundations in Movement Ecology}}
  (\bibinfo{publisher}{Springer Berlin Heidelberg}, \bibinfo{year}{2014}),
  Springer Series in Synergetics, pp. \bibinfo{pages}{177--205}.

\bibitem[{\citenamefont{Benichou et~al.}(2005)\citenamefont{Benichou, Coppey,
  Moreau, Suet, and Voituriez}}]{benichou_optimal_2005}
\bibinfo{author}{\bibfnamefont{O.}~\bibnamefont{Benichou}},
  \bibinfo{author}{\bibfnamefont{M.}~\bibnamefont{Coppey}},
  \bibinfo{author}{\bibfnamefont{M.}~\bibnamefont{Moreau}},
  \bibinfo{author}{\bibfnamefont{P.-H.} \bibnamefont{Suet}}, \bibnamefont{and}
  \bibinfo{author}{\bibfnamefont{R.}~\bibnamefont{Voituriez}},
  \bibinfo{journal}{Physical Review Letters} \textbf{\bibinfo{volume}{94}},
  \bibinfo{pages}{198101} (\bibinfo{year}{2005}), \bibinfo{note}{copyright (C)
  2009 The American Physical Society; Please report any problems to
  prola@aps.org}.

\bibitem[{\citenamefont{Tejedor et~al.}(2012)\citenamefont{Tejedor, Voituriez,
  and Benichou}}]{tejedor_optimizing_2012}
\bibinfo{author}{\bibfnamefont{V.}~\bibnamefont{Tejedor}},
  \bibinfo{author}{\bibfnamefont{R.}~\bibnamefont{Voituriez}},
  \bibnamefont{and} \bibinfo{author}{\bibfnamefont{O.}~\bibnamefont{Benichou}},
  \bibinfo{journal}{Physical Review Letters} \textbf{\bibinfo{volume}{108}},
  \bibinfo{pages}{088103} (\bibinfo{year}{2012}).

\bibitem[{sup()}]{supp_inf}
\bibinfo{note}{URL will be provided by the publisher}.

\bibitem[{\citenamefont{Othmer et~al.}(1988)\citenamefont{Othmer, Dunbar, and
  Alt}}]{othmer_models_1988}
\bibinfo{author}{\bibfnamefont{H.~G.} \bibnamefont{Othmer}},
  \bibinfo{author}{\bibfnamefont{S.}~\bibnamefont{Dunbar}}, \bibnamefont{and}
  \bibinfo{author}{\bibfnamefont{W.}~\bibnamefont{Alt}},
  \bibinfo{journal}{Journal of Mathematical Biology}
  \textbf{\bibinfo{volume}{26}}, \bibinfo{pages}{263} (\bibinfo{year}{1988}).

\bibitem[{Not()}]{Note3}
\bibinfo{note}{$M(x)$ is a monotonously increasing function of its argument and
  has well defined limits: $M(x)\simeq \frac{2}{\sqrt{\pi}}x \log\frac{1}{x}$
  for $x\rightarrow 0$, and $M(x)\simeq 1$ for $x\rightarrow\infty$.}

\bibitem[{\citenamefont{Richardson and
  Lehmann}(2010)}]{richardson_mechanisms_2010}
\bibinfo{author}{\bibfnamefont{B.~E.} \bibnamefont{Richardson}}
  \bibnamefont{and} \bibinfo{author}{\bibfnamefont{R.}~\bibnamefont{Lehmann}},
  \bibinfo{journal}{Nature reviews. Molecular cell biology}
  \textbf{\bibinfo{volume}{11}}, \bibinfo{pages}{37} (\bibinfo{year}{2010}),
  \bibinfo{note}{{PMID:} 20027186}.

\end{thebibliography}

\end{document}